# Design of a common verification board for different back-end electronics options of the JUNO experiment

Y. Yang and B. Clerbaux, Université Libre de Bruxelles, ULB, Brussels, Belgium

*Abstract* — The JUNO observatory is a medium baseline experiment in construction in China. A large liquid scintillator volume detects the antineutrinos issued from nuclear reactors. The liquid scintillator detector is instrumented with 17000 large photomultiplier tubes. Two veto systems are added to reduce the backgrounds. The front-end electronics system performs analog signal processing (the underwater electronics) and after about 100 m cables, the back-end electronics system, outside water, consists of the DAQ and the trigger. One of the main challenges of the whole electronics system is the fast data link (250 Mb/s) combined with the power delivery over 100 m Ethernet cables. Three different options are considered to connect the front-end and the back-end systems, depending on the DAQ data volume and the way to deliver the power to the underwater system. In order to test the three options in an efficient and fast way, a common baseboard with interfaces to different mezzanine boards is designed.

## I. Introduction

The Jiangmen Underground Neutrino Observatory (JUNO) [1] is a neutrino medium baseline experiment in construction in China, with the goal to determine the neutrino mass hierarchy and perform precise measurements of several neutrino mass and mixing parameters, after 6 years of data taking [2,3]. The experiment uses a large liquid scintillator detector aiming at measuring antineutrinos issued from nuclear reactors at a distance of 53 km. The 20 ktons of liquid scintillator contained in a 35 m diameter acrylic sphere is instrumented by more than 17000 20-inch photomultiplier tubes (PMT). Two vetoes are foreseen to reduce the different backgrounds: a 20 ktons ultrapure water Cerenkov pool around the central detector and a muon tracker installed on top of the detector.

Paper submitted on June 24th, 2018.
This work is supported by the F.R.S-FNRS funding agency (Belgium) and by the Université Libre de Bruxelles, ULB.
Yifan Yang is an Ing. Dr. affiliated at the Université Libre de Bruxelles, ULB, Brussels, Belgium (e-mail: Yang.Yifan@ulb.ac.be).
Barbara Clerbaux is a Prof. affiliated to the Université Libre de Bruxelles, ULB, Brussels, Belgium (e-mail: Barbara.Clerbaux@ulb.ac.be).

## II. Readout System

One of the innovative aspects of JUNO is its electronics and readout concept [1,4]. With JUNO, a flexible approach is chosen: each PMT will have its own "intelligence" as self-monitoring (currents, voltages, temperatures) and will have the possibility to implement various data processing algorithms (for example trigger request generation and data compression). The JUNO electronics system can be separated into mainly two parts: (i) the front-end electronics system performing analog signal processing (the underwater electronics) and after about 100 meters cables, (ii) the back-end electronics system, sitting outside water, consisting of the data acquisition (DAQ) and the trigger. The main challenge of the whole electronics system is the very strict criteria on reliability: a maximum of 0.5% failure over 6 years for the PMTs full readout chain, as well as the large data transfer of 250 Mb/s and the power supply that need to be delivered over 100 m Ethernet cables.

Due to the high requirement of reliability, the Ethernet cable, containing 4 twisted differential pairs, is chosen to be the link media between the underwater and the outside water systems. The Ethernet cable has a relatively low price and high robustness.

The information exchanges between the underwater and outside water systems are of 3 types: the trigger related information, the event data and slow control related information, and the power supply. Trigger related signals are concentrated on a TTIM FMC mezzanine card sitting on a Back-End Card (BEC). Event data and slow control related signals are either connected directly to a commercial switch or go first through the BEC and then to the switch. The power supply is provided either through dedicated cables or is combined with the data transfer through the Ethernet cables.

For the front-end electronics, the Global Control Unit (GCU) [4] sitting inside an underwater box digitizes the incoming analog signals with custom designed high speed ADU (analog to digital converter unit). The GCU stores the signals in a large local memory under the control of the FPGA (Field-Programmable Gate Array) waiting for trigger decision, and sends out possible event data as well as trigger requests to the outside-water system. The BEC is used as a concentrator and the incoming trigger request signals pass an equalizer for compensating the attenuation due to the long cables. An FPGA mezzanine card collects all differential trigger request signals, aligns them to a certain trigger count, makes a sum, and sends the result to the trigger system over an optical fiber.

Depending on the different usages of the Ethernet cables and on the power supply method used, 3 possible schemes are proposed for the JUNO readout electronics. They are described below.

### A. The BX scheme

A schematic view of the BX scheme is presented in Figure 1. In this scheme, one GCU unit handles the signals of one PMT, and one Ethernet cable connects one GCU unit to the BEC. The 4 differential pairs inside the Ethernet cable has the following functions:

For the uplink:
1: trigger request signal from the GCU unit to the BEC at 125Mbps
2: event data and slow control feedback over fast Ethernet packet and power on Ethernet (PoE+) DC+
For the downlink:
1: trigger acknowledge signal from the BEC to the GCU unit at 125Mbps and 5W power
2: control command from the BEC to the GCU unit over fast Ethernet packet and PoE+ DC-

On the BEC, the trigger related signals are routed to the TTIM, the event data and slow control signals are routed to the RJ45 connectors and are connected to a commercial PoE+ switch.

over Giga bit Ethernet. Another Ethernet cable connects one GCU unit to the BEC; the 4 differential pairs inside one cable have the following functions:

For the data link:
1: trigger request signal from the GCU unit to the BEC at 250Mbps
2: trigger acknowledge signal from the BEC to the GCU unit at 250Mbps
For the power link:
1: 48 V 15W positive
2: 48 V 15W negative

On the BEC, the trigger related signals are routed to the TTIM, the power links are routed to power connectors and are connected to a commercial power supply.

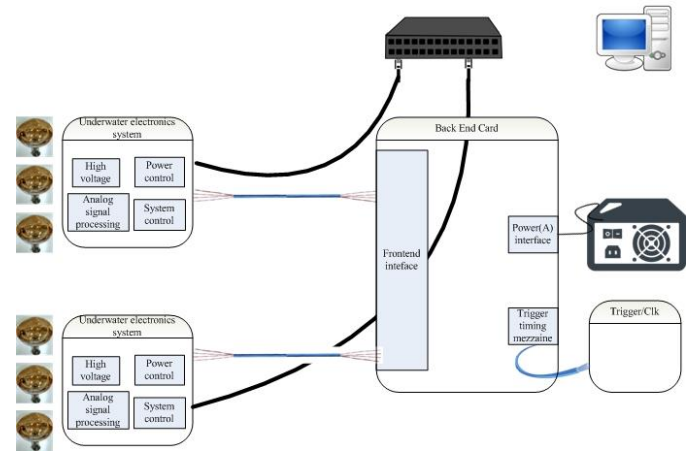

*Figure 2: schematic view of the "1F3 scheme".*

### C. The 1F3RE scheme

In this case, one GCU unit handles 3 PMTs and two Ethernet cables are used to connect the GCU unit to the BEC. A dedicated power cable brings power from commercial power supply to the underwater system. A schematic view of the 1F3RE scheme is presented in Figure 3. The two Ethernet cables are identical, with functions as described below:

For the uplink:
1: trigger request signal from the GCU unit to the BEC at 250 Mbps
2: event data and slow control feedback over fast /Giga bit Ethernet packet
For the downlink:
1: trigger acknowledge signal from the BEC to the GCU at 250Mbps
2: control command from the BEC to the GCU over fast/ Giga bit Ethernet packet

The trigger related signals are routed to the TTIM, the event related signals from the two cables are joined together to one RJ45 connector and are connected to a

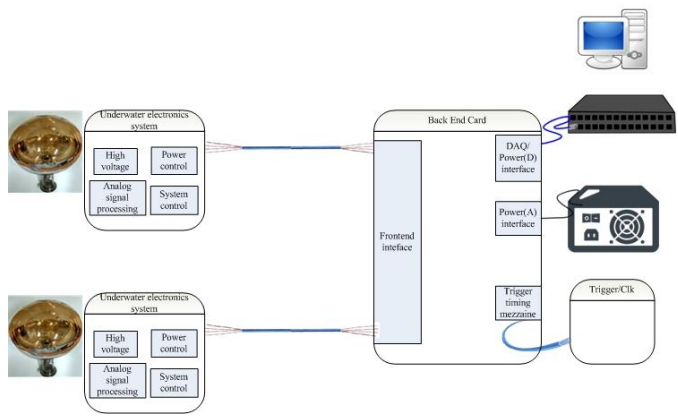

*Figure 1: schematic view of the "BX scheme".*

### B. The 1F3 scheme

In this scheme, presented in Figure 2, one GCU unit handles signals of three PMT. A dedicated Ethernet cable brings power from commercial power over Ethernet switch to the underwater system, it is also in charge of event data and slow control related information exchange commercial Giga bit Ethernet switch. In normal operation mode, two times two event links can make a Giga bit Ethernet link. So in the case of one cable lost, the remaining cable can still operate as a fast Ethernet link with the left two links, and thus provide redundancy.

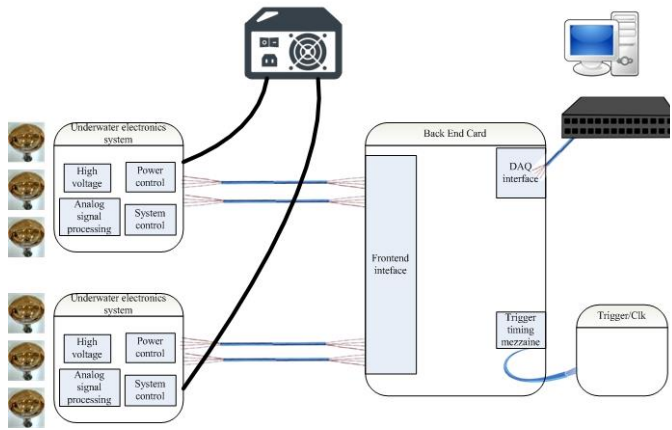

*Figure 3: schematic view of the "*1F3RE *scheme".*

### III. DESIGN

The 3 schemes are presently being studied by the electronics group of the JUNO experiment. A common verification board, which is able to implement different types of tests, is important and critical for the choice of the best scheme and for the qualification of the full data chain. To design a common verification board, the framework of a mother board and a mezzanine card structure is used.

Due to the unavailability of some key components like the GCU unit and the TTIM, the considerations for the design of the verification board are the following:

- Use a baseboard and a mezzanine card structure, the baseboard has the full function of the 1F3 scheme
- Make it possible to test the critical data path without the need of real underwater electronics and real TTIM
- All differential pairs use impedance control including pairs for the power supply
- Each channel for power supply use 12 mils track on 1 oz copper for 1A current with separate resettable fuse protection, 8 channels share one power module
- Power supply for underwater electronics is separated from the power supply for the back-end card, for testing different grounding possibilities
- Implementing individual trigger request and acknowledge connection to the two FMC connectors for maximum flexibility

The baseboard is designed with the stack up configuration as presented in Figure 4.

| Layer | Info | | | Thickness |
|---|---|---|---|---|
| TOP | | | | 0.333+Plating |
| | PP | TU-75P | 106+1080 | 5.086(mil) |
| L2 | | | | 1 Oz |
| | Core | TU-752 | 0.13 | 5.118(mil) |
| L3 | | | | 1 Oz |
| | PP | TU-75P | 1080+2116*2 | 11.856(mil) |
| L4 | | | | 1 Oz |
| | Core | TU-752 | 0.21 | 8.268(mil) |
| L5 | | | | 1 Oz |
| | PP | TU-75P | 1080+2116*2 | 11.883(mil) |
| L6 | | | | 1 Oz |
| | Core | TU-752 | 0.13 | 5.118(mil) |
| L7 | | | | 1 Oz |
| | PP | TU-75P | 106+1080 | 5.074(mil) |
| BOT | | | | 0.333+Plating |

*Figure 4: Stack up chosen for the baseboard.*

The PCB design of the baseboard is presented in Figure 5. The PCB has 48 RJ45 connectors on the left side of the baseboard to provide the connections to the underwater system. Close to the connectors, there are 48 equalizers for the uplink trigger request signals. The 48 output differential pairs connect to 2 custom defined LPC connectors with two serial 0 Ohm resistors in each path. In the middle part of the baseboard, two LPC connectors are used to provide connection to the TTIM. On the right side of the TTIM, another 48 differential pairs connect back to the left RJ45 connectors for the downlink trigger acknowledgement signals. In order to fulfill any future requirement, in total 96 differential pairs are connected to the two LPC connectors.

All the event data and slow control related signals from the RJ45 connectors are routed to the rightest header connectors and then to the power connectors with 0 ohm resistors and with a resettable fuse in between. In the middle of the right part of the baseboard, there is the power connector for the BEC itself, it is separated with the power supply for the underwater system to study flexible grounding schemes.

There are header pin connectors at every key places for the mezzanine card. In total, 5 kinds of mezzanine cards have been developed. There are described below and presented in Figure 6.

- Mezzanine card 1. This first card is an equalizer mezzanine card. In the case of the baseboard equalizers do not function properly, we still have the chance to design a small board with more powerful equalizers to implement the required function without the redesign of all the other parts.

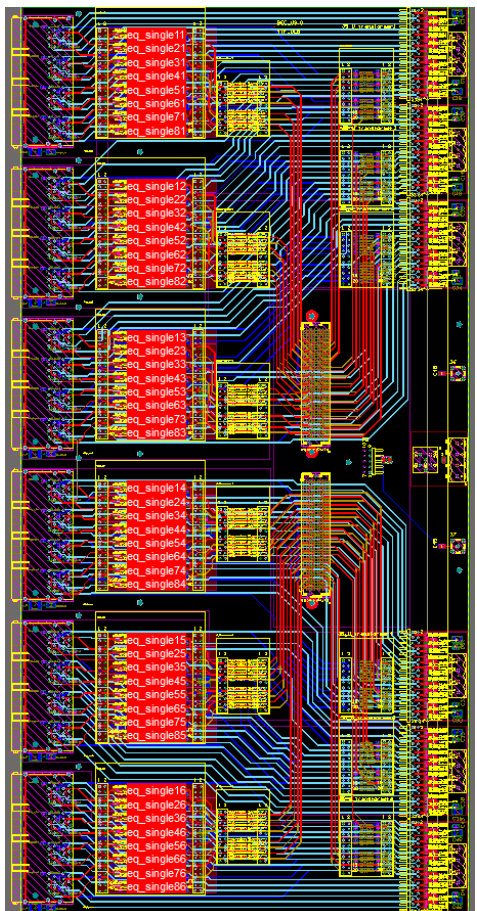

*Figure 5: Design of the BCP of the baseboard.*

- Mezzanine card 2. This is a translator mezzanine card. Since the TTIM is not available, we do not know how many IO are on the TTIM. By using a differential to singe ended translator, we can half the amount of trigger related signals if needed.
- Mezzanine card 3. This is the TTIM emulator, in order to perform the full function tests without the need of the real TTIM board.
- Mezzanine card 4. It is the fan out and cable driver board. The global trigger acknowledge signal is identical for all the channels, thus we can get the 48 channels signals by fanning out a single source, it can decrease the IO requirement for the TTIM design. On the other hand, the trigger acknowledgement signals are delivered to the GCU over 100 m cable. Though they are equalizers on the GCU side, it may still necessary to have cable drivers on the BEC end to increase the transfer capability.
- Mezzanine card 5. It is the header to the RJ45 board. This board is used to combine the event data related signals from two cables to one RJ45 connector, thus making a Giga bit Ethernet link.

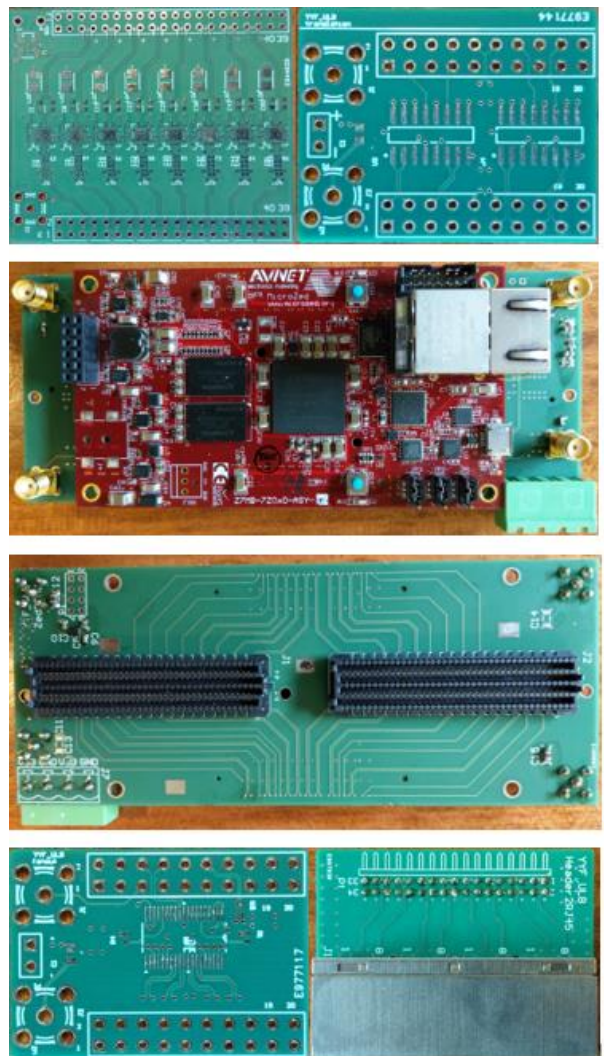

*Figure 6: The 5 mezzanine cards. From top to bottom: the equalizer card; the translator card; the TTIM emulator (top and bottom); the fan out and cable driver card; the RJ45 card.*

IV. TEST

Each BEC has 48 RJ45 connectors and can handle 48 bi-directional trigger links, as well as 48 bi-directional Ethernet link for the BX scheme and the 1F3RE scheme. By connecting two RJ45 connectors with one 100 m cable, we can emulate the GCU unit with one RJ45 port and thus evaluate the performance of the trigger signal transfer as well as the event data transfer. Comparing the 3 schemes, the most demanding requirements are as below:

1: for trigger related signals, 250Mbps bi-directional data transfer.

2: for event and slow control related signals, able to use 4 links combine to a Giga bit Ethernet link.

3: for power related link, able to deliver 15W power.

A schematic description of the general test setup is given in Figure 7, and the test set up for the trigger data transfer is shown in Figure 8.

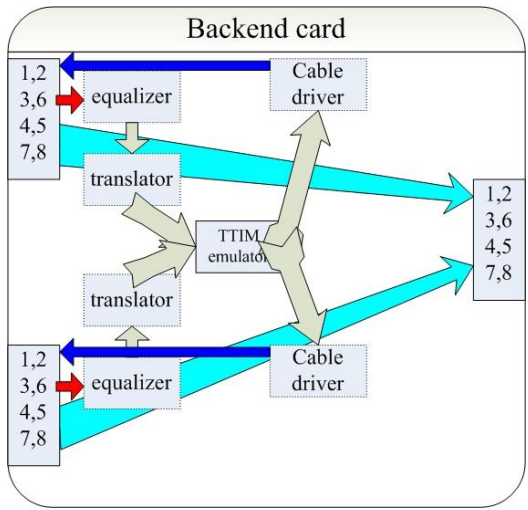

*Figure 7: schematic view of the general test set up.*

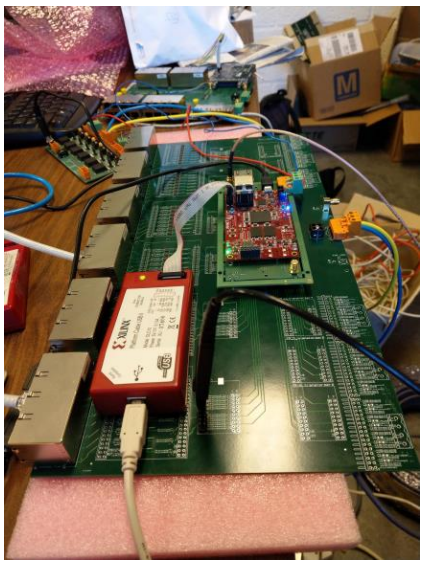

*Figure 8: test setup for the trigger data transfer.*

The test for the trigger data transfer is done as follows. The TTIM emulator generates two 250Mbps prbs-7 data stream. We connect directly to pin 1,2 of two RJ45, to the two other RJ45 connectors, with one 100 m crossover CAT5 cable. We check the eye diagrams on the outputs of two corresponding equalizers (see Figure 9). The results show that the equalizers work properly, indeed they can recover the binary signal after the long cable attenuation, thus two bi-directional 250Mbps data links have been established.

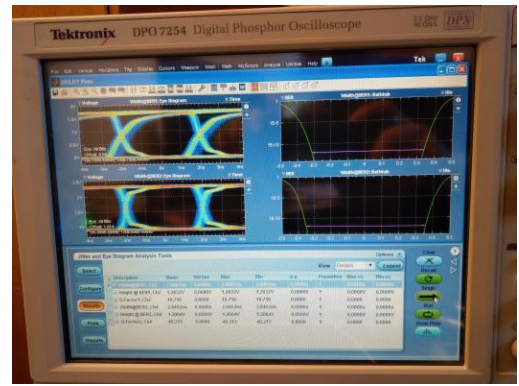

*Figure 9: eye diagrams from the outputs of the two equalizers.*

Another test concerns the Ethernet packet transfer. Due to the lack of a real GCU unit, to avoid using a non-standard Ethernet cable, we can connect one PC to one port of the header to RJ45 board, while we connect another port to a commercial switch. On the left of the BEC, we use two Ethernet cables to connect the corresponding four RJ45 connectors (one RJ45 on the right contains 4 links coming from two Rj45 connectors on the left), thus we established in theory one Giga bit Ethernet link between the PC and the commercial switch. We then run speed test on PC. The result of the test is presented in Figure 10. It shows that the Giga bit Ethernet is achieved. By using this method, we can verify that the routing on the baseboard is fine for both BX and 1F3RE scheme.

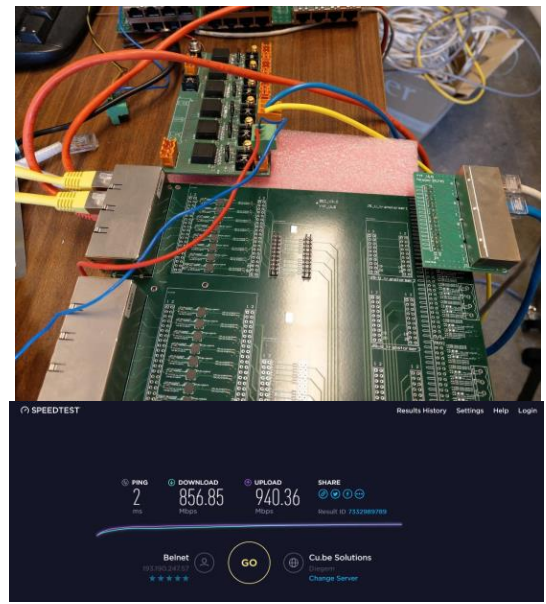

*Figure 10: Ethernet data transfer test.*

A final test concerns the power delivery, the test is relatively easy, on the right side of the baseboard, we connect the power connector for one group of underwater system to a commercial power supply, on the left side of the baseboard, we connect one port inside this group to a 50 W rheostats over 100 m Ethernet cable. By changing the value of the rheostats, we can verify the power delivery capability as well as the function of resettable fuse. The result shows that the baseboard design is able to support a current up to 1A, which means that the 1F3 scheme is fine for the power delivery.

## V. CONCLUSION

We have used a common verification board combined with different mezzanine cards to perform trigger data transfer test as well as Ethernet data transfer and power delivery tests. The results of the tests show that the common verification board is working properly and that the 3 schemes are all feasible.

This common verification board is an important tool in order to provide a critical review of the 3 schemes and finally to choose the best scheme. It is also critical for the qualification of the full data chain. More tests with real JUNO modules are foreseen.